\documentclass{article}
\usepackage[utf8]{inputenc}
\usepackage{latexsym}
\usepackage{float}
\usepackage{graphicx}
\usepackage{amsmath, subfigure}
\usepackage{hyperref}
\usepackage[colorinlistoftodos]{todonotes}
\usepackage{authblk}
\usepackage{lineno}

\newcommand{\nico}[1]{\textcolor{black}{#1}}

\title{\nico{Structured pathways in the turbulence organizing recent oil spill events in the Eastern Mediterranean.}}
\author[1]{Guillermo Garc\'ia-S\'anchez}
\author[1,*]{Ana M. Mancho}
\author[2]{Antonio G. Ramos}
\author[2]{Josep Coca}
\author[3]{Stephen Wiggins}
\affil[1]{Instituto de Ciencias Matem\'aticas, CSIC, C/ Nicol\'as Cabrera 15, Campus Cantoblanco, 28049 Madrid, Spain}
\affil[2]{ Instituto ECOAQUA, Faculty of Marine Sciences. Campus Universitario de Tafira, Universidad de Las Palmas de Gran Canaria, 35017 Las Palmas de Gran Canaria, Spain}
\affil[3]{School of Mathematics, University of Bristol,
Bristol BS8 1TW, United Kingdom}

\date{10th January 2022}

\begin{document}

\maketitle
\abstract{

The chaotic nature of ocean motion is a major challenge that hinders the discovery of spatio-temporal current routes that govern the  transport of material. Certain material, such as oil spills, pose significant environmental threats  and these are enhanced by the fact that they evolve in a chaotic sea, in a way  which still nowadays  is far from being systematically anticipated.
Recently  such an oil spill event has affected the Mediterranean coast of several Middle Eastern countries. No accidents were reported for these spills previous to their arrival at the coast, and therefore there was  no hint of their origin. Modelling such an event, in which uncertainties are increased due to the lack of  information on where and when the spills was produced, stretches available technologies to their limits, and requires the use of novel ideas  that help to understand the essential features of oil and tar transport by ocean currents. In this regard 
Lagrangian Coherent Structures enable us to find order within ocean chaos and provide powerful insights into chaotic events and their  relationships over different locations and times like the one addressed. 
Using the observed locations of  the oil impacting the coast at  specific times, we seek to determine its original location and the time it was released in the open ocean.
We have determined both using a combination of earlier satellite observations and computational modelling of the time evolution.
The observed agreement between modeled cases and satellite observations  highlights the power of these ideas.
  
}

{\bf Introduction. }  At the beginning of  2021 the coastline of several Middle Eastern countries in the Eastern Mediterranean was affected  by the presence of  oil from unknown source(s).  Figure 1 illustrates the affected geographical areas marked with pink and yellow bullets.  A more detailed report on the impacted zones along the Israeli coast may be found in the interactive map whose link is provided in \cite{jp4}. Israeli authorities
estimated that more than 1000 tons of tar \cite{haar} landed along 180 km  of the israeli and lebanese shoreline \cite{jp1,jp2, jp3, ay1,rut,ata} in mid February. 
Gaza also reported that similar arrivals of tar reached their beaches days afterwards \cite{gaza}.  This sequence of events has caused one of the worst ecological disasters reported in decades in the affected countries, which will require years of cooperative action for the affected areas to  be restored \cite{haar}. Numerous photographs and visual material documented the nature of the oil and actions taken by governments to mitigate the disaster \cite{WP, jp1,jp2,jp3,ay1,rut,ata}. The type of tar visible on the images is consistent with a degradation of the released oil for time periods of around a month. 

What was the origin of these spills? Perhaps they could have been  ``deliberate" oil spills, in line with findings by  Pavlakis et al. (1996) \cite{pavlakis96} who report that such oils spills appear with considerably higher frequency than oil spills corresponding to  ship accidents. Also these could have been  caused by operative discharges from ships, since according to the European Space Agency (1998) \cite{ESA98} 45\% of the oil pollution comes from these. There exist systematic efforts to prevent such events. Such as,  for instance, the activities of the CleaSeaNet Service of the European Maritime Safety Agency (EMSA), operating since 2007, to locate and identify polluters in areas under their jurisdiction. EMSA received a request from the Israeli authorities concerning this particular event  \cite{jp3}. 
 
  Is any attempt to reconstruct the sequence of events of this spill with the available information doomed to failure?  Indeed, there exist recent examples in the literature that confirm that predictions on oil spill evolution   still raise big questions. For instance,  in 2015 the fishing ship Oleg Naydenov  caught 
fire and sank in the south of Gran Canaria. There exist models for the spreading of the oil that \nico{however} did not  report neither the date or its arrival point to the coast of Gran Canaria  \cite{ivorra};  in 2018  the Iranian oil tanker Sanchi
collided with a cargo ship, caught 
fire, and sank in the East China Sea. After the event, researchers tried to assess 
where pollutants from the Sanchi would travel, but  there was no consensus between predictions from groups in China and the United Kingdom \cite{sanchi};  recently,  on August 23rd, 2021 a spill from  Syria's largest refinery spread across the Mediterranean. Predictions   expected its arrival to Cyprus  on the 31 August 2021 \cite{siria1}, however by the 6th of September 2021 such a landing had not yet been reported \cite{siria2}. Many other examples could have been added to this list. One aspect that makes the pollutant event addressed in this article 
  particularly challenging is the fact that the date and geographical location where the spill, or the spills, were released are  unknown. That is, if oil spills, such as those just quoted, are difficult to predict even if  the time and position of the event that has produced them are known, the difficulty increases substantially if this  information is not known.

 The first models for oil spill spreading used  simplified linear superposition techniques to model ocean currents that employed a vector sum of the 
 mean  flows, tides, wind/waves and turbulent dispersion  \cite{mit,modeloil}. In contrast, nowadays there exist sophisticated models that integrate
 all these effects to predict  ocean  currents and all relevant ocean variables. These models are run operationally and  provide unprecedented conditions to produce accurate oil spill predictions. 
 One natural question in this context is  if  it is possible to provide answers for the  event under consideration with currently available tools? Among these  are those provided by Copernicus,  one of the most ambitious programmes  in Earth Observation \cite{COP}. Copernicus encompasses 
 the Sentinel programme, which provides very high resolution satellite images in radar and visible frequencies, and the Copernicus Marine Environmental service (CMEMS), which provides data on ocean currents on a  daily basis. 
Despite the availability of these new   products,   oil spill forecasts are  still uncertain  since the underlying ocean flow, and the associated  transport, is very chaotic.
Importing into this setting dynamical systems concepts, which use concepts from chaos theory, may provide a wealth of new ideas that could  assist in this struggle. One of these consist of 
identifying geometrical features on the ocean surface that help to interpret colectively the behaviour of masses of fluid parcels, instead of considering  individual fluid trajectories. 
The development of these geometrical ideas was begun by Poincar\'e in the context of his work on celestial mechanics. In the setting of geophysical flows these geometrical structures have been referred to as Lagrangian Coherent Structures (LCS). For
oil spills this global vision has provided very satisfactory results.  Using this perspective Garcia-Garrido et al \cite{ggrmcw16} were able to identify the date and arrival point of the oil to the coast of Gran Canaria after the Oleg Naydenov fishing ship accident; Garcia Sanchez et al. \cite{garciasanchez2020} used this viewpoint to describe the Volcan Tamasite event and to compare the performance of different ocean models \cite{garciasanchez2022}. It is important to remark that these  episodes occurred in  different space and time scales ranging from the mesoscale,  to the submesoscale, and from hours to days.    Above the mesoscale Olascoaga and Haller \cite{Olas2012} found that oil in the Gulf of Mexico, released after the Deepwater Horizon accident, was lined up with configurations obtained from LCS, which highlighted  attracting material curves. Notably this approach has succeeded for events that involved different type of oils, from light fuel oil in \cite{garciasanchez2020}, to denser IFO 380 oil  \cite{ggrmcw16} and to a variety of crude elements \cite{Olas2012}.
Encouraged by these results we will address the description of the oil spill event described above. 
  With the aid of LCS we identify
  attracting material curves along which oil spills tend to become aligned (see \cite{Olas2012, garciasanchez2022}). Indeed,  this article reports  strong correlations between satellite observations and attracting material curves highlighted by LCS computed with  the CMEMS data. These connections have been very valuable for identifying the sources of the contamination event. The agreement between satellite observations and modelling results supports the correctness of our approach.
 We expect  that these scientific and technological advances will be systematically implemented in the near future and will allow investigators to  identify violators and take appropriate measures to protect the environment. 
 \begin{figure}[htb!]
 \includegraphics[scale=0.8]{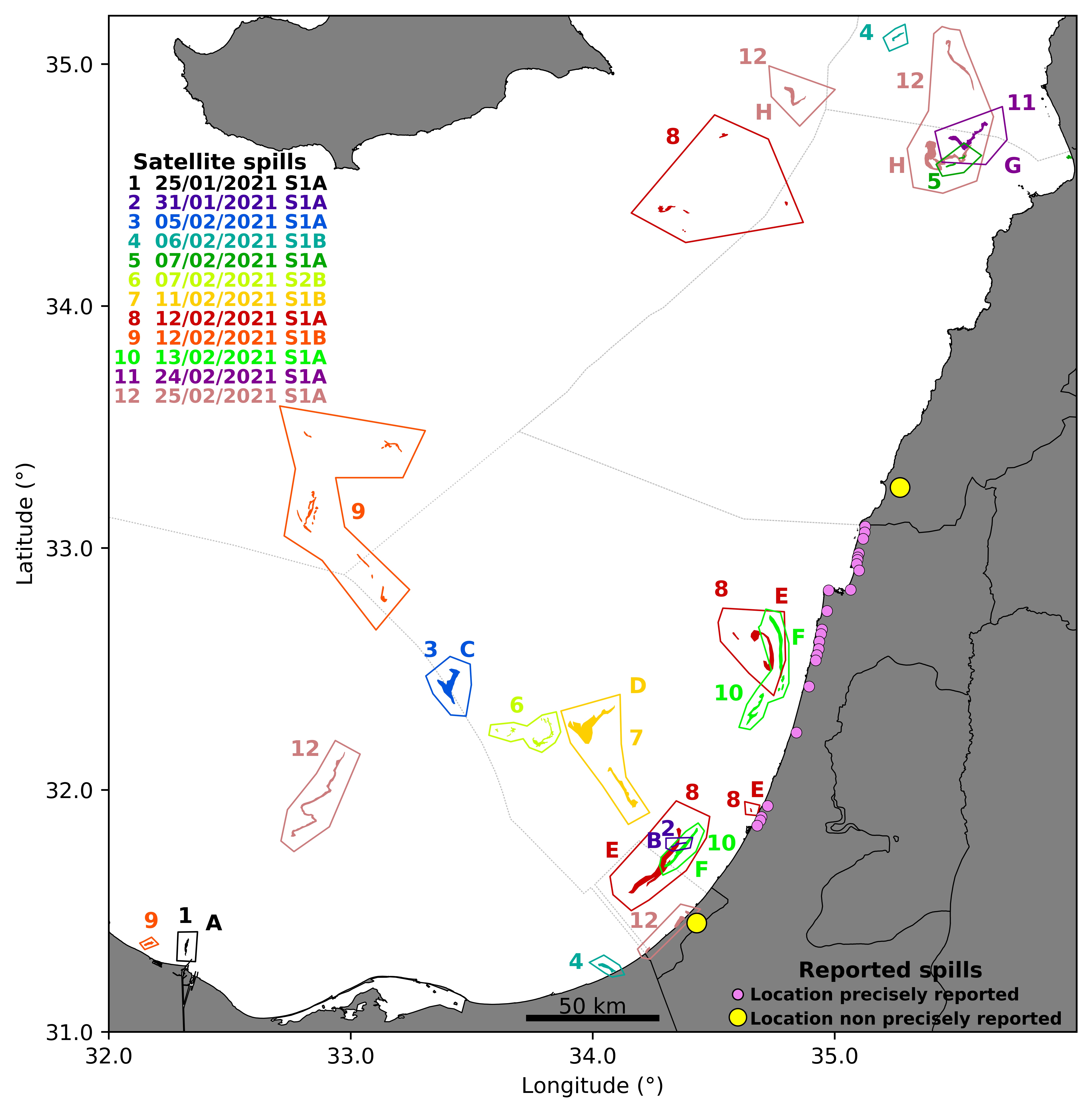}
   \caption{Scenario of  affected areas. Exclusive Economic Zones (EEZ) are marked with a dotted line.  Pink and yellow bullets mark coastal areas for which, according to mass media reports, there were in situ observations of oil or tar. Satellite observations are listed with numbers and colors according to the day in which they were detected. Some of them are double labelled with a letter since they are also reported in figure \ref{fig2}. The figure has been created using python 3.8.10 and matplotlib module 3.1.2 (https://www.python.org/downloads/release/python-3810/). Coastlines and country boundaries have been obtained from GSHHG - A Global Self-consistent, Hierarchical, High-resolution Geography Database 2.3.7 version (https://www.ngdc.noaa.gov/mgg/shorelines/gshhs.html). EEZ have been  obtained from https://www.marineregions.org/. The satellite observations vector representations have been created using QGIS software 3.10 version (https://qgis.org/) and the SAR processed images. }  
   \label{fig1}
 \end{figure}
 

{\bf Satellite Data.} During  the time period and geographic region of interest a set of publicly available satellite data  has been analyzed. The aim was to identify spills on the sea surface that could be linked to the  reported spills along the coastline. 


 One source for these satellite images is the Synthetic Aperture Radar (SAR) high resolution data from Sentinel 1 (A,B), Level-1 IW GRDH (Interferometric Wide Swath Ground Range Detected).  SAR data were  processed  using  the  SNAP  -  ESA  Sentinel  Application  Platform v8.0.3 Graphics Processing Tool (GPT) operators [https://step.esa.int/main/toolboxes/snap/].  During the study period there were days around the third week of February in which, due to the atmospheric conditions, SAR images were not useful. The occurrence of storms, and their associated strong wind fields, produced atmospheric signals that were detected in SAR images. This prevented the information corresponding to surface sea phenomena from being showed. 
For this reason, SAR images interpretation require wind field data, and to support SAR image analysis, wind data with moderate resolution were obtained from Copernicus Marine Segment [https://marine.copernicus.eu/]. The selected product is identified with product code 012\_004V6. This product corresponds to global near real time wind data every 6 hours with a spatial resolution of one-quarter degree. This product combines data from wind models of the ECMWF (European Centre for Medium-Range Weather Forecasts) with data from available wind scatterometers. Additionally, instantaneous wind fields were derived from SAR data using the corresponding GPT  operator.

 Both mineral oil and  biogenic slicks are visible in SAR images at moderate wind speeds. The wind range for which both types of slicks are recognizable is not the same, although they overlap. In order to contradistinguish mineral oil slicks from natural origin surface films, the SAR images were analyzed not only with available wind fields but also with ocean colour data. To this end, 
 Sentinel 2 (A, B) optical data from MultiSpectral Instrument (MSI) Level 1C  and  Sentinel 3 (A,B) Ocean and Land Colour Instrument (OLCI) Level 1B were  downloaded from the Sentinel Data Hub [https://scihub.copernicus.eu/].
Sentinel 2 MSI data atmospheric correction was conducted using\cite{vanhellemont2019} the acolite toolbox [https://odnature.natural sciences.be/remsem/software-and-data/acolite]. Sentinel 3 OLCI data were processed using SeaDAS 7.5.3 version [https://seadas.gsfc.nasa.gov/].

\begin{figure}[htb!]
 \begin{center}
\includegraphics[scale=0.2]{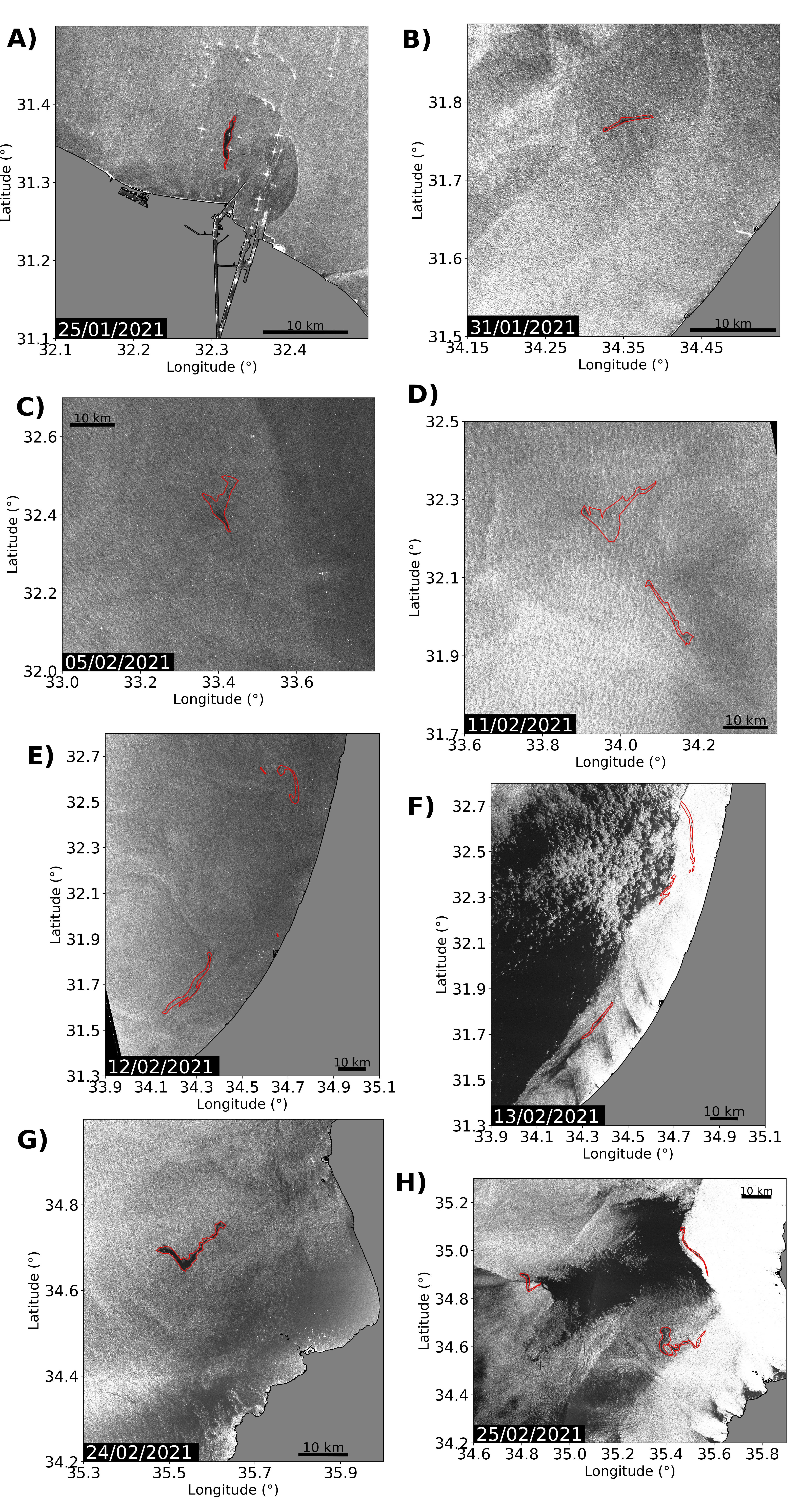} 
\end{center}
   \caption{ Satellite radar  images. a) Sentinel 1A on the 25th January 2021; b) Sentinel 1A on the 31st January 2021; c) Sentinel 1A on the 5th February 2021; d) Sentinel 1B on the 11th February 2021; e) Sentinel 1A on the 12th February 2021; f) Sentinel 1A on the 13th February 2021 ; g) Sentinel 1A on the 24th February 2021; h) Sentinel 1A on the 25th February 2021. The figure has been created using python 3.8.10 and matplotlib module 3.1.2 (https://www.python.org/downloads/release/python-3810/). Coastlines have been obtained from https://osmdata.openstreetmap.de/data/land-polygons.html. The satellite observations vector representations have been created using QGIS software 3.10 version (https://qgis.org/) and the SAR processed images. }
   \label{fig2}
 \end{figure}
 
  {\bf Satellite Results.} Figure \ref{fig2} displays a set of satellite images selected in the area and period of interest after performing  the  analysis described above. This selection is chosen also in the light of the numerical simulations reported next. Panel a) shows a spill detected close to the Suez Canal on the 25th of January 2021. As we will argue in the discussion, this spill has been detected shortly after its release and marks the origin of some of the spills that reached the coastline. Panel b) shows a second spill closer to the coast on the 31st of January 2021. As discussed later we conjecture that this observation is a second origin for those spills that reached the coastline.
 Panel c) displays one spill visible on the 5th of February 2021.
 Panel d) identifies spills on the 11th of February, not too far from the position in c). Panel e) shows spills  very close to the Israeli coast  on the 12th February 2021. Panel f) shows the evolution of these spills  on the 13th February 2021. Panels g) and h) show a sequence of images detected  close to the Syrian coast on the 24th February and 25th February 2021, respectively.

Spots in  Figure \ref{fig2} are displayed by zooming into smaller areas. In order to acquire a vision of their global   distribution they are also represented in Figure \ref{fig1}.  Figure 1 shows the position of all spills located from satellite images in the Eastern Mediterranean during the period of interest.  Each shape appearing in Figure \ref{fig2} is labelled in  Figure \ref{fig1} with the same letter. Figure  \ref{fig1} shows additional spills, some of them reported in the media \cite{greenpeace}, for which we have not found relations to their arrival to the shoreline. These are labelled only with numbers, and their observation day is listed in the figure. Satellite images confirm that waters of the Exclusive Economic Zone (EEZ)  of Egypt,  Syria and Cyprus  were also affected, and not only those of  Israel, Gaza, and Lebanon. In this figure it is remarkable that spills in panels a) and b), which we have identified as the {\it sources},
are seen very small when compared to others. We remark that according to Schrope \cite{schrope2013}  spills are often larger than detected, and their size very often is not correctly determined from remote sensing images. This is particularly true at the early stages when their spreading in the ocean surface has barely started. On the other hand, structures displayed in panels c), d), e), f), g) and h), which are more extended and present distorted shapes, suggest that they have been affected by the chaotic motion of ocean transport.

{\bf The  Transport Problem.} One natural question regarding those oil spills observed by satellite and at the coast is whether  they  are connected among themselves? Are they part of the same event? We  provide answers to these questions in this section. 

Oil  transport can be described by a
concentration field, $C$, whose  evolution  is affected by the flow, represented by the velocity field $\mathbf{v}(\mathbf{x(t)}, t)$, and molecular diffusion. 
At the ocean scales under consideration, the diffusion transport is much smaller than the convective
transport \cite{tesismp} and therefore it can be neglected. On the other hand, particles with a finite size and different density to that of water may not instantly follow fluid velocities: other mechanisms besides passive advection contribute to their
transport, such as gravity forces, their finite size, inertia and
history dependence, etc. Results in \cite{pm1} confirm, however, that for
a wide range of particles, the description of passive tracers is appropriate, except for the
addition of a constant vertical velocity arising from the particle
weight. However, for the case under study, that correspond to oil that remains floating for a long time, weight effects should be negligible and therefore we considered that it
closely follows the 2D horizontal motions of fluid parcels. This is a simplified approach,  but studies such as \cite{lekienpd,Olas2012,ggrmcw16,garciasanchez2020} confirm that considering horizontal advection as
the dominant contribution to pollutant transport provides very
good results. Similar approaches in other contexts of ocean transport confirm the same \cite{ggmwm15}. Finally, we will assume that the crude oil moves horizontally, close to the surface, but a bit below the waterline and therefore it is  not subjected
to direct wind sailing effects. 
Wind effects, jointly with tides, waves, etc., are included in the sophisticated ocean models that provide the velocity fields of the ocean currents, $\mathbf{v}(\mathbf{x(t)}, t)$, (see Section Ocean Data below for more details) and we assume that oil parcels follow instantly water  fluid velocities. The accuracy
of this approach is supported,  by the
agreement between the predictions made by the simulations, the oil
sightings from satellite and in situ observations. Under this approach oil parcels follow trajectories
$\mathbf{x(t)}$ that evolve according to the dynamical system:
\begin{equation}
\frac{d\mathbf{x}}{dt}= \mathbf{v}(\mathbf{x}, t) \label{v}
\end{equation} 
In this equation the position {\bf x} is described in longitude ($\lambda$) and latitude ($\phi$)
coordinates, that is, {\bf x} = $(\lambda, \phi)$, and {\bf v} represents the velocity
field. In longitude/latitude coordinates, the dynamical system \eqref{v}
can be rewritten as:
\begin{equation}
\frac{d \lambda}{dt}= \frac{u(\lambda,\phi,t)}{R \cos(\phi)}, \,\,\,\,\,\,\,\,\, \frac{d \phi}{dt}= \frac{v(\lambda,\phi,t)}{R}, \label{lanlot}
\end{equation} 
where $R$ is the Earth's radius. This system assumes that the
vertical velocity component in the ocean is small compared to
the horizontal ones and for that reason it has been disregarded.
The two velocity components are determined by the zonal (u) and
meridional (v) velocities, which are obtained from the currents provided by CMEMS.

There exist diverse software packages that are able to track
oil spills\nico{, of which for instance a list is found  in} \cite{garciasanchez2020}. Most of these models are focused on tracking
individual fluid parcels, and in order to maintain a good representation
of the spill they  play with a sufficiently
large number of initial parcels. Contrary to these approaches, in the results reported in this work,
we track in time the whole area where the fuel is extended,
and the algorithm self regulates the number of fluid parcels on
the contour to ensure its accurate representation at all times.
 At the beginning the area is a simple shape, but while it evolves, it becomes distorted  and  convoluted. The shape is tracked with contour advection algorithms developed by \cite{dri}, including some modifications explained in\cite{physd,npg1,physrep}.  Typical oil spill software packages  include models to represent the weathering process according to different oil properties. In our approach we do not describe  oil transformation. We consider that weathering does not affect to the transport process.   Spill contours are evolved uncoupled from degrading effects. Degrading effects could have been considered
at representative levels {\em a posterior} \cite{garciasanchez2020} as a change in the  color intensity of the oil
spill. It could be that weathering implies oil evaporation, and in that case  the evolved contour would be an upper bound to the oil evolution. Similarly would occur if oil forms clots that are denser and  sink. In both cases  the region of dispersion for the oil is a subset of our approach. Given that we do not have information on the type of floating  oil  and our focus is just on transport.

{\bf Ocean Data.}
The ocean velocity fields used in this work were obtained from the CMEMS [http://marine. copernicus.eu/]. In particular, we have used
the datasets provided by the high resolution Global Ocean Model (the global analysis
and forecast product).  The horizontal resolution of the model is 1/12$^\circ$ (approximately 8 km) with regular longitude/latitude equirrectangular projection and 50 vertical geopotential levels.  Data are served on a daily and hourly basis.  Hourly data  rapidly fluctuate around daily data, which represent smooth averaged values. All the calculations reported in this article have been performed  with the daily data, because they better match observations.  
Additionally, CMEMS provides a specific service with hourly and daily data for the Mediterranean area. The resolution of the Mediterranean Model is higher than that of the Global Model: 
on the horizontal plane it is 1/24$^\circ$ with regular longitude/latitude equirrectangular projection and along the vertical coordinate it has 141 depth levels.
This  resolution allows a representation of sub-mesoscale features, however   calculations 
performed with the Mediterranean Model do not reproduce many of the transport features related to the oil spill event. Additionally we have found that small scale features from this Model do not accurately correspond to those observed from satellites.

{\bf The Dynamical Systems Perspective.}
A challenge to studying Eq. \eqref{v} is that even flows with smooth velocity fields may exhibit complex particle trajectories. An
approach taken from nonlinear dynamical systems theory, seeks 
to understand the behavior of large ensembles of particle trajectories by finding geometrical structures, known as Lagrangian Coherent Structures (LCS), that form time dependent material surfaces. This perspective has been successfully employed in pollution contexts \cite{lek, Olas2012, ggrmcw16, garciasanchez2020}. The LCS spatio-temporal template  can be constructed with a recent technique referred to as Lagrangian Descriptors
(LDs). The particular LD that we use is a function referred to as $M$ \cite{madrid2009,mendoza2010,mancho2013} which is defined as follows:
\begin{eqnarray}
   && M({\bf x}_0,t_0,\tau) = \int_{t_0-\tau}^{t_0+\tau} \|{\bf v}({\bf x}(t),t)\|\ dt = \nonumber\\
    && \int_{t_0}^{t_0+\tau} \|{\bf v}({\bf x}(t),t)\|\ dt +\int_{t_0-\tau}^{t_0} \|{\bf v}({\bf x}(t),t)\|\ dt \,,
    \label{M}
\end{eqnarray}
Singular features visible in this field represent attracting  and repelling   material surfaces. Of particular interest for our study are attracting material surfaces, visible from the backward integration (the second term)  of Eq. \eqref{M}, along which oil blobs  eventually tend to be elongated and aligned. \nico{More specifically, there exist regions in the ocean, characterized by high contraction and expansion rates, referred to as hyperbolic regions, such that  blobs placed in the neighbourhood of trajectories in these regions (hyperbolic trajectories)  evolve in time  rapidly expanding and filamenting to become aligned with the attracting material curves. This effect has direct implications on our study since blobs that are originally located in  small ocean regions, if they go close to these expansive/contractive regions,  will spread in filaments affecting  large sea areas. Alternatively, repelling material curves are obtained from the forward integration (the first term)  of Eq. \eqref{M}. These curves describe  how  blobs placed in the neighbourhood of the hyperbolic region evolve in backward time. They also tend to form filaments, but as these are formed  in reverse time, they are not observable and for this reason they are referred to as repelling material curves.  A physical way to look into these curves is as follows:  material spread on the ocean surface lined up with these filaments evolves in time contracting towards the  neighbourhood  of the hyperbolic trajectory.}

The computation of fluid particle trajectories {\bf x}$(t)$ is necessary in order to
evaluate the function  $M$ in Equation \eqref{M}. For a given initial condition  {\bf x}$_0(t)$ this function evaluates  the arc length of trajectories when they are evolved forwards and backwards in time
for a period $\tau$. Trajectories are integrated with a 5$^{th}$ order Runge-Kutta method, and arc length is computed by the addition of linear segments connecting successive steps of the Runge-Kutta method.

\begin{figure}[htb!]
 \begin{center}
  \includegraphics[scale=0.31]{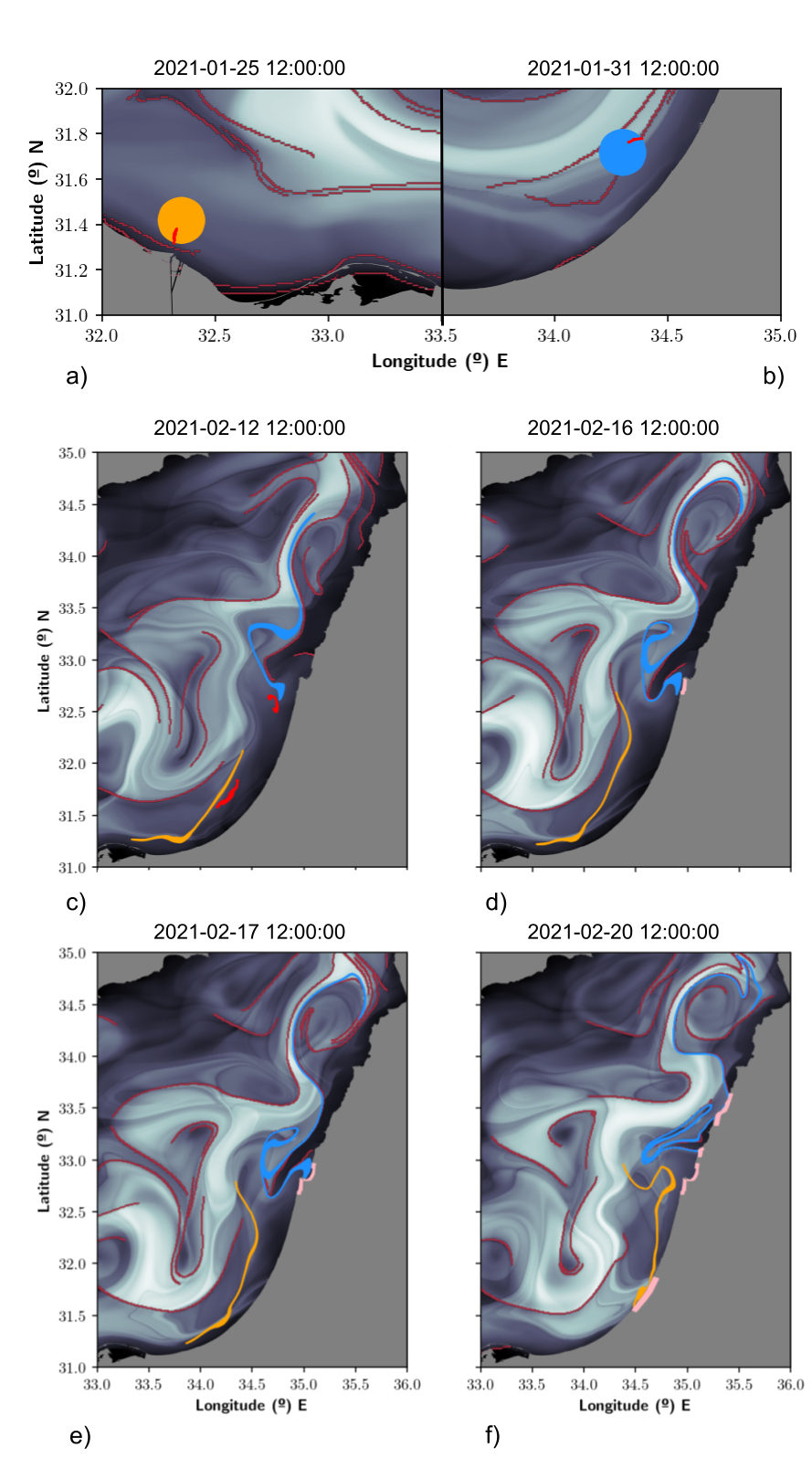}
\end{center}
   \caption{Simulations that reproduce the spreading on the sea and the arrival of the oil to the coast. The function $M$ at the background is represented at  $t_0$ equal to the corresponding date.  a) 1st release on the 25th January 2021 and satellite observation; b) 2nd release on the 31st January 2021 and satellite observation; c) evolution on the 12th February 2021 and satellite observations; d) evolution on the 16th February 2021 and arrival to the coast; e) evolution on the 17th February 2021 and arrival to the coast; f) evolution on the 25th February 2021 and arrival to the coast. These masks indicate regions that correspond to the continental shelf.   These figures have been created with Python 3.9.2 (https://www.python.org/downloads/release/python-392/). The maps shown have been generated with a mask provided by OpenStreetMaps (https://osmdata.openstreetmap.de/data/land-polygons.html).}
   \label{fig3}
 \end{figure}

{\bf Transport Results.} Figure  \ref{fig3} represents in gray tones the field $M$ as evaluated from  Eq. \eqref{M} with $\tau=$ 15 days, which is a choice appropriate for the time scale of the described events (1 month). In maroon tones are highlighted the attracting material curves.  Repelling material lines are also visible, but we do not highlight them as they are not of interest for our discussion \nico{at this moment}. This figure represents the time evolution of orange and blue blobs that have been released on the 25th and 31st of January, respectively, at the positions marked in panels a) and b), which are linked to spills detected in panels a) and b) of Figure    \ref{fig2}. Initial blobs have a radius of 4 km, in agreement with the resolution of the CMEMS global model. It is clear from the evolution that blobs tend to end up elongated and aligned with the maroon features of the attracting material curves.
Panel c) illustrates the evolution of these blobs on the 12th of February and the red shapes highlight satellite spills  spotted on that day. The good agreement between parts of the evolved orange and blue blobs and these spills is remarkable. Panels d) and e) illustrate  their evolution on the 16th and 17th of February and the arrival to the Israeli coast at  points marked in pink and also reported in Figure \ref{fig1}. Panel f) illustrates the arrival at further points on the Israeli and Lebanon coasts   on the 20th of February, in good agreement with reports marked in  Figure \ref{fig1}.

\begin{figure}[htb!]
 \begin{center}
  \includegraphics[scale=.2]{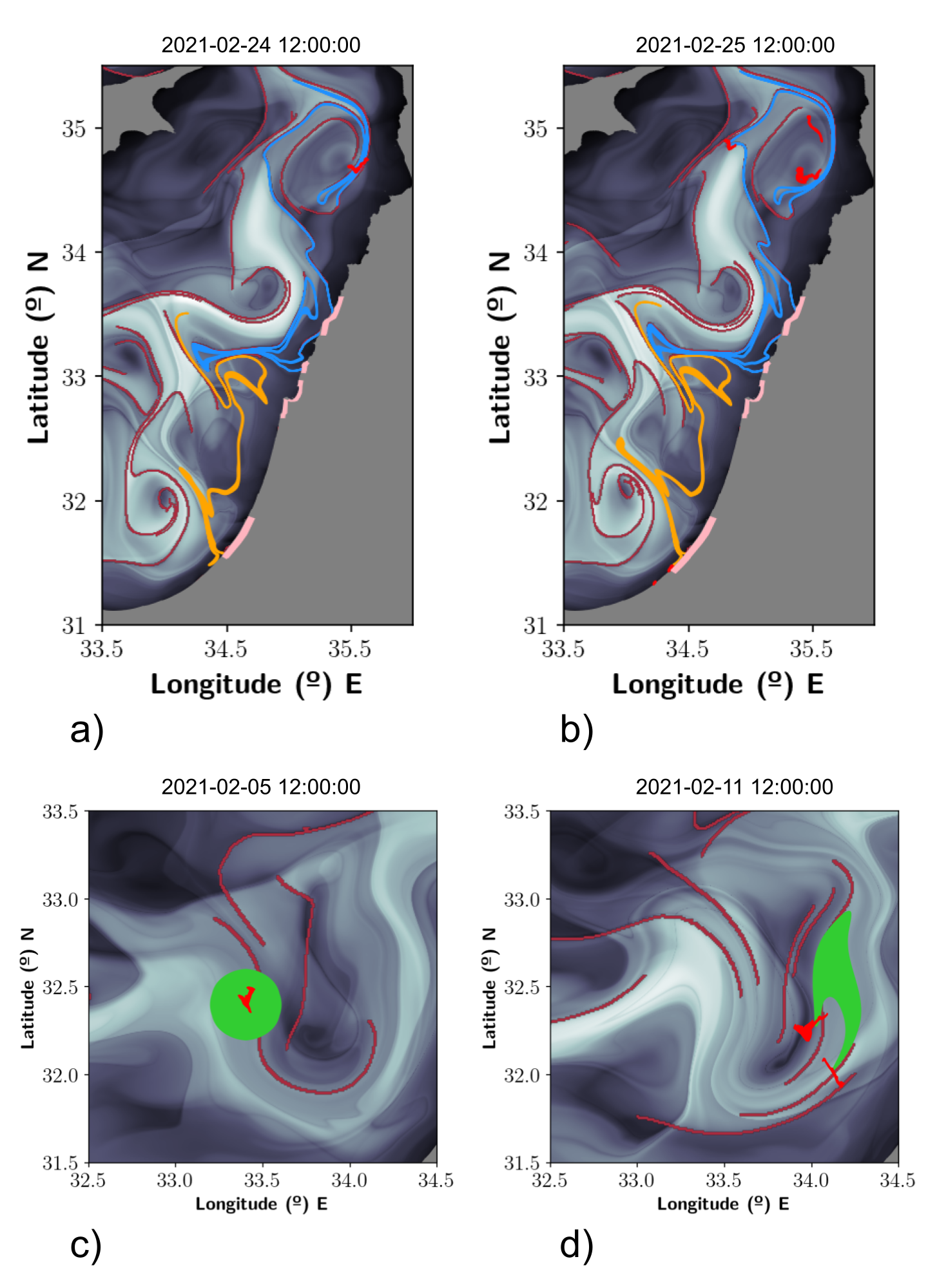}
\end{center}
   \caption{Simulations that connect successive oil spill observations. The function $M$ at the background is represented at  $t_0$ equal to the corresponding date. a) Evolution on the 24th February 2021 and satellite observations; b) evolution on the 25th February 2021 and satellite observations; c) satellite observation on the 5th February 2021 and a 1st release attached to the observation; d) evolution on the 11th February 2021 and satellite observations.  These figures have been created with Python 3.9.2 (https://www.python.org/downloads/release/python-392/). The maps shown have been generated with a mask provided by OpenStreetMaps (https://osmdata.openstreetmap.de/data/land-polygons.html).}
   \label{fig4}
 \end{figure}

Figure  \ref{fig4} is similar to Figure \ref{fig3}. It also represents in gray tones the field $M$ and the attracting material curves   highlighted in maroon. Panels a) and b) complete the description of
Figure \ref{fig3}.
Panel a) illustrates the blue and orange blobs on the 24th of February. Their structure is extremely filamented, closely following the attracting material curves. The red shape illustrates the spill visible in  panel   \ref{fig2}g) suggesting that this could have been related to the spill detected in  panel   \ref{fig2}b), associated with the blue blob. 
Panel   \ref{fig2}h) confirms  the connection between the "V" shaped spill in g) with the one observed in h) confirming the quality of the model \eqref{v} to provide a consistent connection of this spill with what is obtained from satellite observations. In this panel  it is also remarkable the proximity between all satellite observations (in red), the blue blob and the attracting material curves.   This suggests that these spots observed on the 25th of February close to the Syrian coast could also been related to  the spill detected in  Figure    \ref{fig2}b).  Finally on the Gaza coast  are visible two red spots that highlight satellite observations of oil. There is a remarkable agreement between these observations and the spreading of the orange blob. 

Panels c) and d) illustrate the position,  respectively,  of spills detected in panels c) and d) of Figure    \ref{fig2}. The green blob in panel \ref{fig3}c) is released at the position of the spill of panel   \ref{fig2}c) on the 5th of February  and its evolution on the 11th of February according to the model \eqref{v}  is illustrated in panel \ref{fig3}d). It is remarkable the connections established by the simulations between both satellite observations, suggesting that these spills correspond to the same event. Also we want to emphasize that these spills are not related to spills arriving to the coast.
Arrivals to the coast  are only achieved by the blue and orange blobs visible in Figure \ref{fig3} and in panels a) and b) previously discussed.
The movie supplied in the supplementary material completes the description given in Figures \ref{fig2} and \ref{fig3}.

In all panels of figure  \ref{fig3} there exist  V-shaped observed satellite spills. We conjecture that these shapes are again an effect of advection dominated transport. Indeed the convoluted forms adopted by  blobs while being transported,  visible in Figures \ref{fig2} and \ref{fig3}, present many  corners with this kind of shape, at  different orientations. Even the green blob
in panels  \ref{fig3}c) and d) is distorted into a V-shape.
"V" shapes are a footprint of typical stretching and folding mechanisms related to hyperbolicity and non-linearity present in equations like \eqref{v}, adopted to describe oil transport\cite{ottino1994reynolds}.

\nico{Finally,  we discuss the perspective that searches for contamination sources by considering backwards integration of trajectories in the spirit of \cite{shu1, shu2}.  Figure S1 of the supplementary information shows these results. Panel a) of this figure locates a blob in the neighbourhood of the Israeli coast on the date that tar was reported to reach the beach, the 16th February 2021. Panel b) in this figure, represents the backward evolution of this blob on the  31st January 2021. As anticipated, the backward evolution of the original blob becomes aligned with the repelling material curves that are highlighted in blue in the figure and spread over a large ocean region. This points out the difficulty in locating the spill point through this methodology. }

{\bf Discussion and Conclusions.} 
This paper discusses  our findings concerning a recent oil spill event that we have studied by  importing dynamical systems ideas. We have found evidence that our perspective recovers the essential features of the oil spill dispersion at large scale.
Indeed oil spills form clots  and its pieces  tend to be aligned with segments of attracting material curves, Lagrangian Coherent Structures, confirming the assumptions.
Indeed, the attracting material curves of the CMEMS Global model are closely related to SAR satellite spill observations for this Eastern Mediterranean event. Remarkably, several of these observed scattered spills emerge from two particular observations on the 25th and 31st January 2021 and are directly linked to the coastal spill arrivals.  We also find that some observations of spills on the 5th and 11th of February 2021 visible from  Sentinel SAR images  are unrelated to spills observed at coastal points, however they seem to be connected between themselves. The entire description of the event, according to our findings,  spans a period of almost one month and this is consistent with the fact that what has been reported to reach the coast is tar, a form of degraded oil after several weeks on the sea surface.

Connections and links between scattered pieces of oil observed from satellites and on coastal arrivals have been established for the CMEMS global model.  However these types of connections are not found from other CMEMS models such as that in the Mediterranean sea domain.  This suggest that CMEMS global model is particularly good for describing transport phenomena. 

Dynamical system tools have provided concepts such as that of attracting material curves which have provided a simple global overview of the event and suggested connection routes for dispersed and scattered spills,  helping to answer questions about  where and when the spill originated.  It is remarkable that these connections have been established for events that extend for a one month period, taking oil spill forecast capacities to their limit\cite{barker2020}.   

As a conclusion we have found that the synergistic combination of Copernicus services has provided a powerful technology that should be exploited in an operational manner to better predict and target the evolution at sea of  spill events.  These technologies will allow a better environmental protection of all seas and coasts, given that, as in this case, oil spills pay no attention to international borders.


{\bf Acknowledgements.}
The authors thank E.U. Copernicus Marine Service Information for providing wind fields data. Authors thank the SeaDAS Development Group at NASA GSFC for providing SeaDAS software. The authors thank E.U. Copernicus Sentinel Data for providing Sentinel 1,2 and 3 data used in the present paper. The authors thank REMSEM - OD Nature Remote Sensing and Ecosystem Modelling team for providing acolite software.
The authors acknowledge support from IMPRESSIVE, a project
funded by the European Union's Horizon 2020 research and
innovation programme under grant agreement No 821922, which main goal is to better predict and targeting the evolution at sea of potential spills events with emergent observing satellite and predicting techniques based on the dynamical systems theory. 
SW acknowledges
the support of ONR Grant No. N00014-01-1-0769. AMM acknowledges the support of CSIC PIE project Ref. 202250E001. AMM is an active member of  two CSIC Interdisciplinary Thematic Platforms: POLARCSIC and TELEDETECT.

 \bibliographystyle{plain}

\end{document}